\documentclass[english]{article}
\usepackage[T1]{fontenc}
\usepackage[latin1]{inputenc}
\usepackage{amsmath}
\usepackage{amssymb}

\makeatletter

\let\SF@@footnote\footnote
\def\footnote{\ifx\protect\@typeset@protect
    \expandafter\SF@@footnote
  \else
    \expandafter\SF@gobble@opt
  \fi
}
\expandafter\def\csname SF@gobble@opt \endcsname{\@ifnextchar[%]
  \SF@gobble@twobracket
  \@gobble
}
\edef\SF@gobble@opt{\noexpand\protect
  \expandafter\noexpand\csname SF@gobble@opt \endcsname}
\def\SF@gobble@twobracket[#1]#2{}

\usepackage{babel}
\makeatother
\begin{document}

\newcommand{\bra}[1]{\langle#1}

\newcommand{\ket}[1]{|#1\rangle}

\newcommand{\F}{\Phi^{0}}

\newcommand{\s}{\bar{\Sigma}}

\newcommand{\f}{\Phi}

\newcommand{\Om}{\Omega}

\newcommand{\Si}{\Sigma}

\newcommand{\om}{\Omega^{0}}

\newcommand{\tdot}{\cdot\cdot\cdot}

\part*{The Presentation of the Quantum Algebra of Observables of the Closed
Bosonic String in 1+3 Dimensions: The Presentation in Manifestly Lorentz
Covariant Form%
\footnote{ This work is based on the author´s diploma thesis \cite{D.P04}%
}}

\hspace{0,5cm}

\begin{center} D. Peter

Physikalisches Institut, Albert-Ludwigs-Universität,

Hermann-Herder-Str. 3, D-79104 Freiburg i. Br., Germany\end{center}

\hspace{1cm}

Abstract: The quantum algebra of observables of the massive closed
bosonic string in 1+3 dimensions has been developed so far in the
rest frame of the string. In this paper a method to write this algebra
in a manifestly Lorentz covariant form is explained and compared with
an alternative approach in the literature.

\subsection*{1. Introduction}

The quantum observables of the Nambu-Goto theory form a $\mathbb{N}$-graded
algebra generated by reparametrization invariant quantities $\hat{{\cal Z}}_{\mu_{1}\tdot\mu_{N}}^{(K)}$.
Here $N$ is rank of the tensor whereas $K$ denotes the grade of
$homogenity$. To distinguish between quantized and classical objects
the former will be denoted by a hat. The grading of the algebra is
given by the number $l=N-K-1$, called `degree`, corresponding to
a physical dimension $\mathsf{h}^{l+1}:=\left(\frac{\hbar}{2\pi\alpha^{\prime}}\right)^{l+1}$.
Here $\alpha^{\prime}$ denotes the inverse string tension. Classically
the generators ${\cal Z}_{\mu_{1}\tdot\mu_{N}}^{(K)}$ form a Poisson
algebra. Without loss of generality, the investigation is restricted
to the part $\hat{\mathfrak{h}}^{-}$ of the algebra, the part of
the $\mathit{rightmovers}$. Moreover, the string is thought to be
massive and moving in 3+1 dimensional Minkowski space with metric
(+,-,-,-).

$\hat{\frak{h}}^{-}$ carries a representation of $O(3)$. So its
elements can be organized in multiplets of spin and parity denoted
by $J^{P}$. The spin of a given object of the algebra will be indicated
by a subscript. In Ref.\cite{KP99} the quantum algebra spin $j$
multiplets corresponding to the commutator ({[},{]}) or anticommutator
(\{,\}) of two elements of the quantum algebra as a new element of
the algebra are denoted by $\left[,\right]_{j}$ and $\left\{ ,\right\} _{j}$
respectively. It is assumed that the reader is familiar with \cite{KP99},
the notation of which is being used here. In \cite{KP99} an algorithm
for the presentation both of the classical and the quantum algebra
is described. It starts from the spin operator $\hat{{\cal {\cal J}}}_{1}$
of order $\mathsf{h}$ and of the elements $\hat{{\cal S}}_{1}$,$\hat{{\cal {\cal S}}}_{2}$,
$\hat{{\cal {\cal T}}}_{2}$ and $\hat{{\cal B}}_{0}^{(1)}$ of power
$\mathsf{h}^{2}$ and subsequently cycle by cycle by introducing new
elements $\hat{{\cal {\cal B}}}_{0}^{(2l+1)}$ of the subalgebra $\hat{\frak{a}}$
in every order $\mathsf{h}^{2l+2},l=0,1,...$, the generating relations
are determined. $\hat{{\cal J}}_{1}$, $\hat{{\cal S}}_{1}$, $\hat{{\cal S}}_{2}$,
$\hat{{\cal T}}_{2}$ and $\hat{{\cal B}}_{0}^{(2l+1)}$have spin
and parity: $1^{+},1^{-},2^{-},2^{+},0^{+}$ respectively. On the
classical level the former generate a Poisson subalgebra $\frak{U}$
of $\frak{h}^{-}$ whereas the infinite set of elements ${\cal {\cal B}}_{0}^{(2l+1)}$
forms an abelian subalgebra $\frak{a}$ such that $\frak{h}^{-}=\frak{a}\ltimes\frak{U}$. 

The consistency of this algorithm has been proved pending on the verification
of certain hypotheses \cite{CM}.

The semidirect structure of the algebra is only true to leading (classical)
order. In the quantum case a relation corresponding to the order $\mathsf{h}^{6}$
has been found which mixes $\hat{\frak{a}}$ and $\hat{\frak{U}}$
(see \cite{GH02}).

All these attempts have been made in the rest frame of the string
because there the algebra is considerably simplified. G.Handrich was
the first to make the attempt to write the quantum algebra in a manifestly
Lorentz covariant form (ref. \cite{GH02b}).

\subsection*{2. Covariance via a tetrad formalism}

In \cite{GH02b} the manifest Lorentz covariance is achieved via a
tetrad formalism. The tetrad $e_{\underline{\nu}}^{\mu}$, where the
underlined index $\nu$ labels the constituing vectors (0,...,3) and
$\mu$ denotes the component of the respective vector, is defined
as follows: $e_{\underline{0}}^{\mu}=\frac{1}{m}P^{\mu}$, the other
three vectors are orthogonal to $e_{\underline{0}}^{\mu}$ and orthonormalized.
A specific choice would be%
\footnote{Note that the $\delta_{i}^{\mu}$ in the original paper has the wrong
sign.%
}:

\[
e_{\underline{i}}^{\mu}=\left\{ \begin{array}{c}
\frac{1}{m}P_{i}\hspace{2cm}\mu=0\\
-\delta_{i}^{\mu}+\frac{P^{\mu}P_{i}}{m(P_{0}+m)}\hspace{1cm}\mu\in\{1,2,3\}\end{array}\right.i\in\{1,2,3\}.\]
The spacelike vectors have three $SO(3)$ gauge degrees of freedom. 

To transform a tensor from one inertial frame to another, one has
to contract all its components with the tetrad vectors. Furtheron
one has to take care that the $SO(3)$ spin corresponding to the gauge
degree introduced here has to be zero for physical objects. The further
calculations are performed in the spinor representation of the Lorentz
group. Therefore every tensor is labelled by three spin indices $J,K,L$
i.e. $T=T_{J}^{K,L}$ . The two upper indices belong to the representation
of the Lorentz group, the lower index denotes the spin regarding the
$SO(3)$ gauge degree. So the tetrad can be written as:

$e_{0}\rightarrow e_{0}^{\frac{1}{2}\frac{1}{,2}}$ and $e_{1},e_{2},e_{3}\rightarrow e_{1}^{\frac{1}{2},\frac{1}{2}}.$

To perform the formalism it is convenient to define the $\mathit{standard}\, polynomials$:

\[
E_{J}^{K,L}:=\left(\left(\left(e_{1}^{\frac{1}{2},\frac{1}{2}}\right)^{J}\right)_{J}^{\frac{J}{2},\frac{J}{2}}\left(\left(e_{0}^{\frac{1}{2},\frac{1}{2}}\right)^{r}\right)_{0}^{\frac{r}{2},\frac{r}{2}}\right)_{J}^{L,K},r=2\max(L,K)-J.\]

The only generator of degree zero is the Pauli-Lubanski vector $w$:

\[
\hat{w}:=m\sqrt{3}\left(e_{1}\hat{J}_{1}\right)_{0}^{\frac{1}{2},\frac{1}{2}}.\]

The manifest covariant generators of degree one are defined as the
irreducible components of the following covariant tensor:

\[
Q_{\mu\nu\rho\sigma}=\frac{1}{6}\varepsilon_{\mu\nu}^{\hspace{0,25cm}\alpha\beta}\varepsilon_{\rho\sigma}^{\hspace{0,25cm}\gamma\delta}{\cal Z}_{\alpha\beta\gamma\delta}^{(2)}\]
with the totally antisymmetric tensor $\varepsilon^{\mu\nu\alpha\beta}$
where $\varepsilon^{0123}=1.$

The generators $R_{0}^{0,0},$ $R_{0}^{1,1}$ are the Ricci parts
and $W_{0}^{2,0}$, $W_{0}^{0,2}$the Weyl parts of the tensor $Q_{\mu\nu\rho\sigma}$.
The manifestly covariant generators can be expressed as a linear combination
of the rest frame generators and vice versa. 

However the covariant generators are not entirely unique. There exists
for example the following identity:

\[
\left(\hat{{\cal J}}_{1}^{2}\right)_{0}=\frac{2}{\sqrt{3}}\left(\hat{w}\hat{w}\right)_{0}^{0,0}=\frac{1}{16\sqrt{3}}\hat{R}_{0}^{0,0}+\frac{\sqrt{3}}{8}\left(E_{0}^{1,1}\hat{R}_{0}^{1,1}\right)_{0}^{0,0}.\]

The way to pass from the relations in the rest frame to the covariant
relations will be discussed in one typical example of a covariant
relation of degree two: Since the covariant generators are linear
combinations of the rest frame ones, one has to build the following
linear combination of several rest frame relations to produce a relation
between the covariant generators \cite{GH02b}:

\[
0=\left\lfloor -24\sqrt{5}\left(E_{2}^{1,1}\left(\left[\hat{{\cal B}}_{0}^{(1)},{\cal \hat{{\cal T}}}_{2}\right]_{2}-\sqrt{6}i\left[\hat{{\cal S}}_{2},\hat{{\cal S}}_{1}\right]_{2}+a\left(\mathsf{h}\right)^{2}\hat{{\cal T}}_{2}+b\left(\mathsf{h}\right)^{2}(\hat{{\cal J}}_{1}^{2})_{2}\right)\right)_{0}^{1,1}\right.\]

\[
-8\sqrt{3}\left(E_{1}^{1,1}\left(\left[\hat{{\cal {\cal B}}}_{0}^{(1)},\hat{{\cal S}}_{1}\right]_{1}+6\sqrt{\frac{2}{5}}i\left[\hat{{\cal T}}_{2},\hat{{\cal S}}_{2}\right]_{1}-2\sqrt{\frac{3}{5}}\left[\hat{{\cal T}}_{2},\hat{{\cal S}}_{1}\right]_{1}-12\sqrt{\frac{2}{5}}i\left[\left(\hat{{\cal J}}_{1}^{2}\right)_{2},\hat{{\cal S}}_{2}\right]_{1}\right.\right.\]

\[
\left.-\left.\left.12\sqrt{\frac{3}{5}}\left[\left(\hat{{\cal J}}_{1}^{2}\right)_{2},\hat{{\cal S}}_{1}\right]_{1}+d\left(\mathsf{h}\right)^{2}\hat{{\cal S}}_{1}\right)\right)_{0}^{1,1}\right\rfloor \]

\[
-\frac{112}{\sqrt{5}}\left(E_{2}^{1,1}\left[\left(\hat{{\cal J}}_{1}^{2}\right)_{2},{\cal \hat{{\cal B}}}_{0}^{(1)}\right]_{2}\right)_{0}^{1,1}\]

\[
+448\sqrt{6}i\left(E_{2}^{1,1}\left(\left[\left(\hat{{\cal J}}_{1}^{2}\right)_{2},\hat{{\cal T}}_{2}\right]_{2}+\frac{\sqrt{7}}{2}\left[\left(\hat{{\cal J}}_{1}^{2}\right)_{0},\hat{{\cal T}}_{2}\right]_{2}\right)\right)_{0}^{1,1}\]

\[
-192\sqrt{\frac{3}{5}}\left(E_{2}^{1,1}\left[\left(\hat{{\cal J}}_{1}^{2}\right)_{2},\left(\hat{{\cal J}}_{1}^{2}\right)_{0}\right]_{2}\right)_{0}^{1,1}\]

\[
+64\sqrt{6}i\left(E_{1}^{1,1}\left(\left[\left(\hat{{\cal J}}_{1}^{2}\right)_{2},\hat{{\cal S}}_{1}\right]_{1}+\frac{\sqrt{5}}{2}\left[\left(\hat{{\cal J}}_{1}^{2}\right)_{0},\hat{{\cal S}}_{1}\right]_{1}\right)\right)_{0}^{1,1}\]

\[
-\frac{64}{\sqrt{3}}\left(E_{0}^{1,1}\left[\left(\hat{{\cal J}}_{1}^{2}\right)_{0},{\cal \hat{{\cal B}}}_{0}^{(1)}\right]_{0}\right)_{0}^{1,1}\]

\[
-\frac{256}{\sqrt{5}}\left(E_{0}^{1,1}\left[\left(\hat{{\cal J}}_{1}^{2}\right)_{2},\hat{{\cal T}}_{2}\right]_{0}\right)_{0}^{1,1}.\]
$a,b,d$ are open parameters, which may be fixed by the consistency
requirements in higher degrees of the algebra.

In the $\left\lfloor \right.$ $\left.\right\rfloor $ brackets there
are two generating relations of degree two, the other relations (one
in each line) are induced by using $\hat{{\cal J}}_{1}$. The combination
of the rest frame generators into covariant ones yields:

\[
\left[\hat{R}_{0}^{0,0},\hat{R}_{0}^{1,1}\right]_{0}^{1,1}+2\sqrt{\frac{3}{5}}\left[\hat{R}_{0}^{1,1},\left(\hat{W}_{0}^{2,0}+\hat{W}_{0}^{0,2}\right)\right]_{0}^{1,1}+\left(\mathsf{h}\right)^{2}\frac{2a-b+4d}{6}\hat{R}_{0}^{1,1}\]

\[
+\left(\mathsf{h}\right)^{2}\frac{-2a+b+4d}{2}\left(E_{0}^{1,1}\hat{R}_{0}^{1,1}\right)_{0}^{1,1}+\left(\mathsf{h}\right)^{2}\sqrt{\frac{3}{5}}(2a+b)\left(E_{0}^{1,1}\left(\hat{W}_{0}^{2,0}+\hat{W}_{0}^{0,2}\right)\right)_{0}^{1,1}\]

\[
+\left(\mathsf{h}\right)^{2}\frac{2a-b-20d}{3}\left(E_{0}^{2,2}\hat{R}_{0}^{1,1}\right)_{0}^{1,1}=0.\]

This simple example shows very clearly the problems of this approach
regarding the presentation of the algebra.

The main point of criticism is the fact that one has to use both generating
and induced relations to produce covariant relations. The important
thing here is that the relations give information about the algebra:
the generating ones contain new information of the degree in which
they are introduced, the induced ones contain information already
known from earlier degrees. If these relations are mixed one cannot
be sure about the quality of information in the covariant relation.

Another point is of rather technical (but still important) matter:
In the degrees of the algebra there exist only very few (< 10) generating
relations but a largely increasing number of induced relations ($\sim100,000$
in the fifth degree). Although the induced relations are of little
importance for the algebra in the rest frame, they are necessary for
the tetrad formalism. Looking at the number of them one can imagine
that the numerical work can hardly be done.

Furthermore, as has been shown, there exist additional relations between
the generators.

For these reasons a different approach shall be presented now.

\section*{3. Covariance via the Energy-Momentum Vector}

The method to be presented here heavily relies on the use of the energy-momentum
vector of the string. The starting point is again the covariant tensor
$Q_{\alpha\beta\gamma\delta}$. But now it is split into different
generators of the algebra via partial projection on the energy-momentum
vector or its dual. These generators are the irreducible parts of
the following tensors:

\[
\tilde{\Omega}^{\alpha\beta}=\frac{1}{8m^{2}}\varepsilon^{\alpha\kappa\mu\rho}\varepsilon^{\beta\lambda\gamma\sigma}P_{\kappa}P_{\lambda}Q_{\mu\rho\gamma\sigma},\]

\[
\tilde{\Sigma}^{\alpha\beta}=\frac{1}{2m^{2}}\varepsilon^{\alpha\gamma\sigma\rho}g^{\beta\nu}P^{\mu}P_{\gamma}Q_{\sigma\rho\nu\mu}.\]

The irreducible parts are:

\[
\Omega^{\alpha\beta}:=\tilde{\Omega}^{\alpha\beta}+\frac{1}{3}\left(g^{\alpha\beta}-\frac{P^{\alpha}}{m}\frac{P^{\beta}}{m}\right)g^{\gamma\sigma}\tilde{\Omega}_{\gamma\sigma}\hspace{1,9cm}\left(\negthickspace\mbox{ containing all elements of }T_{2}\!\right),\]
\[
\mbox{Tr}\left(\tilde{\Omega}\right):=g^{\gamma\sigma}\tilde{\Omega}_{\gamma\sigma}\hspace{7,2cm}\left(\negthickspace\mbox{ containing }{\cal B}_{0}^{(1)}\!\right),\]

\[
\Sigma^{\alpha\beta}:=\frac{1}{2}(\tilde{\Sigma}^{\alpha\beta}+\tilde{\Sigma}^{\beta\alpha})\hspace{4,8cm}\left(\negthickspace\mbox{ containing all elements of }S_{2}\!\right),\]

\[
\bar{\Sigma}^{\alpha}:=\frac{1}{4}\frac{P_{\gamma}}{m}\varepsilon^{\gamma\alpha\sigma\rho}(\tilde{\Sigma}_{\sigma\rho}-\tilde{\Sigma}_{\rho\sigma})\hspace{3,75cm}\left(\negthickspace\mbox{ containing all elements of }S_{1}\!\right).\]
In the rest frame the covariant generators reduce to the original
generators. 

In addition, one makes use of the tensor \[
\tilde{\Phi}^{\alpha\beta}=\frac{1}{16m^{2}}g^{\alpha\mu}g^{\beta\nu}P^{\kappa}P^{\lambda}Q_{\mu\kappa\nu\lambda}\]
whose irreducible parts are:\[
\Phi^{\alpha\beta}:=\tilde{\Phi}^{\alpha\beta}+\frac{1}{3}\left(g^{\alpha\beta}-\frac{P^{\alpha}}{m}\frac{P^{\beta}}{m}\right)g^{\gamma\sigma}\tilde{\Phi}_{\gamma\sigma}\hspace{1cm}\left(\negthickspace\mbox{ containing all elements of }\left(J_{1}^{2}\right)_{2}\!\right)\]
and

\[
\negthickspace\mbox{ Tr}\left(\tilde{\Phi}\right):=g^{\gamma\sigma}\tilde{\Phi}_{\gamma\sigma}\hspace{6,6cm}\left(\negthickspace\mbox{ containing }\left(J_{1}^{2}\right)_{0}\!\right).\]
These are of course not new generators but are introduced because
they are often needed. To calculate all elements of the algebra including
the relations, one has to know how the covariant generators are coupled
by the commutators. For this, an important feature of the covariant
generators is that in the rest frame all non totally spacelike components
vanish i.e. that the generators look as follows:

\[
\left(\begin{array}{cc}
0 & \begin{array}{ccc}
0 & 0 & 0\end{array}\\
\begin{array}{c}
0\\
0\\
0\end{array} & \left(\begin{array}{ccc}
\cdot & \cdot & \cdot\\
\cdot & \cdot & \cdot\\
\cdot & \cdot & \cdot\end{array}\right)\end{array}\right)\mbox{ or }\left(\begin{array}{c}
0\\
\cdot\\
\cdot\\
\cdot\end{array}\right)\hspace{1cm}(x).\]

The key point in solving the problem is that a $\mathit{real}$ Lorentz
covariant tensor with the property ($x$) is in the same class regarding
symmetry and traces as its spacelike subtensor in the rest frame.

Therefore it is sufficient to investigate the symmetries of the tensors
containing the generators of the algebra in the rest frame with respect
to the $SO(3)$ and show that the coupling procedure does not alter
the property ($x$) of the tensor. 

It is a well known fact that the symmetric and traceless tensors of
rank $J$ carry the irreducible representation of $SO(3)$. Furtheron
it is well known that the coupling of two objects with spin $J_{1}$
and $J_{2}$ yields the Clebsch-Gordan-series so the output is a linear
combination of symmetric and traceless tensors of the ranks $|J_{1}-J_{2}|$
to $J_{1}+J_{2}$. Explicitly one does the coupling of two tensors
of spin $J_{1}$ and $J_{2}$ to spin $J$ with $J_{1}+J_{2}-J$ even
via $\frac{1}{2}\left(J_{1}+J_{2}-J\right)$ traces by turns over
one index of the first and one of the second tensor. If $J_{1}+J_{2}-J$
is odd, one takes $\frac{1}{2}\left(J_{1}+J_{2}-J-1\right)$ traces
by turns over one index of the first and one of the second tensor
and finally contracts two of the remaining indices, one of each tensor,
in which the object obtained so far is antisymmetric, via an $\varepsilon^{ijk}\mbox{ where }\varepsilon^{123}=1$
to one index. How this works and what the prefactors are shall be
shown now.

Regarding $SO(3)$, the tensors to begin with can all be chosen symmetric
and traceless. Therefore the number of indices of each tensor is equal
to the spin $J$ of the tensor. Coupling two of those tensors with
$J_{1}$ and $J_{2}$ indices leads to a sum of irreducible representations.
The procedure will be done in the angular momentum basis with basis
vectors $\mathbf{e}_{+},\mathbf{e}_{3},\mathbf{e}_{-}$:$\,\mathbf{e}_{\pm}=\frac{1}{\sqrt{2}}\left(\mathbf{e}_{1}+i\mathbf{e}_{2}\right).$

Theorem 1: The tensors belonging to the rest frame spin $J$ have
the following form:

\[
\ket{J;M}=\sqrt{\left(\begin{array}{c}
2J\\
J+M\end{array}\right)}^{-1}\]

\[
\times\left\{ \begin{array}{c}
\sum_{r=0}^{min\left(\frac{J+M}{2},\frac{J-M}{2}\right)}(-)^{\frac{J+M-2r}{2}}\sqrt{2}^{2r}S\left(\Omega^{\left(\frac{J+M-2r}{2}\right),(2r),\left(\frac{J-M-2r}{2}\right)}\right),\\
\sum_{r=0}^{min\left(\frac{J+M-1}{2},\frac{J-M-1}{2}\right)}(-)^{\frac{J+M-2r-1}{2}}\sqrt{2}^{2r+1}S\left(\Omega^{\left(\frac{J+M-2r-1}{2}\right),(2r+1),\left(\frac{J-M-2r-1}{2}\right)}\right),\end{array}\right.\]

where, with regard to the indices, the numbers in the brackets denote
the number of plus-, 3- and minusvalued indices respectively. From
now on this notation shall be used unless mentioned otherwise. The
symbol $S$ stands for the symmetrisation of the tensor. The upper
line after the bracket stands for the case $J+M$ even, the lower
one for the other case. This form can be proved easily via induction.
Note that these tensors are all symmetric and traceless because they
carry irreducible representations of the $SO(3)$.

Theorem 2: The coupling of two tensors $A$ and $B$ of spin $J_{1}$
and $J_{2}$ respectively gives the following result:

\[
\left[A^{J_{1}},B^{J_{2}}\right]_{M}^{J}=(i)^{J_{1}+J_{2}+J}\bra{J_{1},J_{2};J_{1},J-J_{1}}\ket{J;J}\sqrt{\left(\begin{array}{c}
2J_{2}\\
J_{2}+J-J_{1}\end{array}\right)}\]

\[
\times\Pi\left(\mu_{\frac{1}{2}(J_{1}+J_{2}-J)+1}\nu_{\frac{1}{2}(J_{1}+J_{2}-J)+1}\cdot\cdot\cdot\mu_{J_{1}}\nu_{J_{2}}\right)\]

\vspace{0,2cm}

\[
\times g_{\mu_{1}\nu_{1}}\cdot\cdot\cdot g_{\mu_{\frac{1}{2}(J_{1}+J_{2}-J)}\nu_{\frac{1}{2}(J_{1}+J_{2}-J)}}\left[A^{\mu_{1}\cdot\cdot\cdot\mu_{J_{1}}},B^{\nu_{1}\cdot\cdot\cdot\nu_{J_{2}}}\right]\]
for $J_{1}+J_{2}-J$ even, and

\[
\left[A^{J_{1}},B^{J_{2}}\right]_{M}^{J}=(i)^{J_{1}+J_{2}+J}\frac{1}{\sqrt{2}}\bra{J_{1},J_{2};J_{1},J-J_{1}}\ket{J;J}\sqrt{\left(\begin{array}{c}
2J_{2}\\
J_{2}+J-J_{1}\end{array}\right)}\]

\[
\times\Pi\left(\mu_{\frac{1}{2}(J_{1}+J_{2}-J-1)+1}\nu_{\frac{1}{2}(J_{1}+J_{2}-J-1)+1}\cdot\cdot\cdot\mu_{J_{1}-1}\nu_{J_{2}-1},\alpha\right)\]

\vspace{0,2cm}

\[
\times g_{\mu_{1}\nu_{1}}\cdot\cdot\cdot g_{\mu_{\frac{1}{2}(J_{1}+J_{2}-J-1)}\nu_{\frac{1}{2}(J_{1}+J_{2}-J-1)}}\frac{P^{\nu}}{m}\varepsilon_{\nu\hspace{0,1cm}\mu_{J_{1}}\nu_{J_{2}}}^{\hspace{0,1cm}\alpha}\left[A^{\mu_{1}\cdot\cdot\cdot\mu_{J_{1}}},B^{\nu_{1}\cdot\cdot\cdot\nu_{J_{2}}}\right]\]
in the other case, where $\bra{J_{1},J_{2};J_{1},J-J_{1}}\ket{J;J}$
denotes the Clebsch-Gordan coefficient of the coupling $J_{1},J_{2}\rightarrow J$
with $M_{1}=J_{1}$ and $M=J$, and $\Pi(\mu_{1}...\mu_{n})$ the
projector on the space of the totally symmetric traceless tensors.
The explicit form of $\Pi(\mu_{1}...\mu_{n})$ is given in the appendix.

The proof shall be only sketched because the calculations are rather
tedious. We will couple two tensors with spin $J_{1}$ and $J_{2}$
to spin $J$ and maximal weight $M=J$. All other weights can be achieved
by the use of the lowering operator. Furthermore we take $J_{1}+J_{2}-J$
to be even, the other case is almost identical though additional care
is advised.

First one makes use of the fact that the tensors are symmetric and
traceless. Therefore every component $\ket{J;M}$ of the tensor can
be written as:

\[
\ket{J;M}=\sqrt{\left(\begin{array}{c}
2J\\
J+M\end{array}\right)}\left\{ \begin{array}{c}
(-)^{\frac{J+M}{2}}\Omega^{\left(\frac{J+M}{2}\right),(0),\left(\frac{J-M}{2}\right)}\hspace{1,5cm}\mbox{{ for J+M even}}\\
(-)^{\frac{J+M-1}{2}}\frac{1}{\sqrt{2}}\Omega^{\left(\frac{J+M-1}{2}\right),(1),\left(\frac{J-M-1}{2}\right)}\,\mbox{{for J+M odd}}\end{array}\right..\]

In the coupling procedure one gets a sum over all possible numbers
$M_{1}$ with $M_{1}+M_{2}=J$. One has to split this sum into those
terms where $J_{1}+M_{1}$ is even and where it is odd. Then the explicit
expression for the Clebsch- Gordan-coefficients and the identities:

\[
\left(\begin{array}{c}
2n\\
2k\end{array}\right)=\sum_{r=0}^{n-k}2^{2r}\left(\begin{array}{c}
n-2r\\
k-r\end{array}\right)\left(\begin{array}{c}
n\\
2r\end{array}\right)\mbox{ and }\]

\[
\left(\begin{array}{c}
2n\\
2k+1\end{array}\right)=\sum_{r=0}^{n-k-1}2^{2r+1}\left(\begin{array}{c}
n-2r-1\\
k-r\end{array}\right)\left(\begin{array}{c}
n\\
2r+1\end{array}\right)\]

are used which yield the following expression for the rhs of $\ket{J;M}$:

\[
(-)^{\frac{J+J_{1}+J_{2}}{2}}\bra{J_{1},J_{2};J_{1},J-J_{1}}\ket{J;J}\sqrt{\left(\begin{array}{c}
2J_{2}\\
J_{2}+J-J_{1}\end{array}\right)}\]

\[
\times\left(\sum_{M_{1}}\sum_{r=0}^{\frac{J_{1}-M_{1}}{2}}2^{2r}\left(\begin{array}{c}
\frac{J_{1}+J_{2}-J}{2}-2r\\
\frac{M_{1}+J_{2}-J}{2}-r\end{array}\right)\left(\begin{array}{c}
\frac{J_{1}+J_{2}-J}{2}\\
2r\end{array}\right)\right.\]

\[
\times\left[A^{\left(\frac{J_{1}+M_{1}}{2}\right),(0),\left(\frac{J_{1}-M_{1}}{2}\right)}B^{\left(\frac{J_{2}+J-M_{1}}{2}\right),(0),\left(\frac{J_{2}-J+M_{1}}{2}\right)}\right]\]

\vspace{0,2cm}

\[
+\sum_{M_{1}}\sum_{r=0}^{\frac{J_{1}-M_{1}-1}{2}}(-)^{\frac{J+J_{1}+J_{2}}{2}}2^{2r}\left(\begin{array}{c}
\frac{J_{1}+J_{2}-J}{2}-2r\\
\frac{M_{1}+J_{2}-J-1}{2}-r\end{array}\right)\left(\begin{array}{c}
\frac{J_{1}+J_{2}-J}{2}\\
2r\end{array}\right)\]

\[
\times\left.\left[A^{\left(\frac{J_{1}+M_{1}-1}{2}\right),(1),\left(\frac{J_{1}-M_{1}-1}{2}\right)}B^{\left(\frac{J_{2}+J-M_{1}-1}{2}\right),(1),\left(\frac{J_{2}-J+M_{1}-1}{2}\right)}\right]\right).\]

Here the first sum extends over all possible $M_{1}$ with $J_{1}-M_{1}$
even and the third sum over all $M_{1}$ with $J_{1}-M_{1}$ odd.
Now every component 

\[
A^{\left(\frac{J_{1}+M_{1}}{2}\right),(0),\left(\frac{J_{1}-M_{1}}{2}\right)}B^{\left(\frac{J_{2}+J-M_{1}}{2}\right),(0),\left(\frac{J_{2}-J+M_{1}}{2}\right)}\]

is split up in the sum via tracelessness and symmetry into \[
A^{\left(\frac{J_{1}+M_{1}-2r}{2}\right),(2r),\left(\frac{J_{1}-M_{1}-2r}{2}\right)}B^{\left(\frac{J_{2}+J-M_{1}-2r}{2}\right),(2r),\left(\frac{J_{2}-J+M_{1}-2r}{2}\right)}.\]

Reordering these indices in a different way, where a definite number
of indices with the value (+) are separated via a colon and the remaining
ones are ordered in the standard way, yields the following result:

\[
A^{\left(\frac{J_{1}-J_{2}+J}{2}\right);\left(\frac{J_{2}-J+M_{1}-2r}{2}\right),(2r),\left(\frac{J_{1}-M_{1}-2r}{2}\right)}B^{\left(\frac{J_{1}-M_{1}-2r}{2}\right),(2r),\left(\frac{J_{2}-J+M_{1}-2r}{2}\right);\left(\frac{J_{2}-J_{1}+J}{2}\right)}.\]
Here it is clear that the sum over the inner three sets of indices
can be performed so that only the indices at the extreme positions
remain. 

After the transformation into a cartesian basis the final result is
:

\[
\left[A^{J_{1}},B^{J_{2}}\right]_{M}^{J}=(i)^{J_{1}+J_{2}+J}\bra{J_{1},J_{2};J_{1},J-J_{1}}\ket{J;J}\sqrt{\left(\begin{array}{c}
2J_{2}\\
J_{2}+J-J_{1}\end{array}\right)}\]

\[
\times\Pi\left(\mu_{\frac{1}{2}(J_{1}+J_{2}-J)+1}\nu_{\frac{1}{2}(J_{1}+J_{2}-J)+1}\cdot\cdot\cdot\mu_{J_{1}}\nu_{J_{2}}\right)\]

\vspace{0,2cm}

\[
\times g_{\mu_{1}\nu_{1}}\cdot\cdot\cdot g_{\mu_{\frac{1}{2}(J_{1}+J_{2}-J)}\nu_{\frac{1}{2}(J_{1}+J_{2}-J)}}\left[A^{\mu_{1}\cdot\cdot\cdot\mu_{J_{1}}},B^{\nu_{1}\cdot\cdot\cdot\nu_{J_{2}}}\right].\]

To do this in a manifestly covariant manner one has to replace $\delta^{ij}$
for the traces by $-g^{\mu\nu}$ and $\varepsilon^{ijk}$ by $\frac{P_{\mu}}{m}\varepsilon^{\mu\nu\alpha\beta}$.
One easily verifies that this replacement does not change the property
$(x)$ of the objects. Note that the specific form of the prefactors
arises from the choice of the component of the spin $J$ multiplet
with highest weight. The value is of course independent of this choice.
In order to illustrate the method more explicitly and to compare it
with the tetrad formalism, we present one example to demonstrate the
transformation of a typical relation of degree two from rest frame
notation into manifestly covariant form. For convenience the following
definitions are introduced: \[
\Pi\left(\mu_{\frac{1}{2}(J_{1}+J_{2}-J)+1}\nu_{\frac{1}{2}(J_{1}+J_{2}-J)+1}\cdot\cdot\cdot\mu_{J_{1}}\nu_{J_{2}}\right)g_{\mu_{1}\nu_{1}}\cdot\cdot\cdot g_{\mu_{\frac{1}{2}(J_{1}+J_{2}-J)}\nu_{\frac{1}{2}(J_{1}+J_{2}-J)}}\times\]

\[
\left[A^{\mu_{1}\cdot\cdot\cdot\mu_{J_{1}}},B^{\nu_{1}\cdot\cdot\cdot\nu_{J_{2}}}\right]:=[A,B]^{\mu_{\frac{1}{2}(J_{1}+J_{2}-J)}\nu_{\frac{1}{2}(J_{1}+J_{2}-J)}\cdot\cdot\cdot\mu_{J_{1}}\nu_{J_{2}}}\]
and 

\[
\Pi\left(\mu_{\frac{1}{2}(J_{1}+J_{2}-J-1)+1}\nu_{\frac{1}{2}(J_{1}+J_{2}-J-1)+1}\cdot\cdot\cdot\mu_{J_{1}-1}\nu_{J_{2}-1},\alpha\right)\times\]

\[
g_{\mu_{1}\nu_{1}}\cdot\cdot\cdot g_{\mu_{\frac{1}{2}(J_{1}+J_{2}-J-1)}\nu_{\frac{1}{2}(J_{1}+J_{2}-J-1)}}\frac{P^{\nu}}{m}\varepsilon_{\nu\hspace{0,1cm}\mu_{J_{1}}\nu_{J_{2}}}^{\hspace{0,1cm}\alpha}\left[A^{\mu_{1}\cdot\cdot\cdot\mu_{J_{1}}},B^{\nu_{1}\cdot\cdot\cdot\nu_{J_{2}}}\right]\]

\[
:=[A,B]^{\mu_{\frac{1}{2}(J_{1}+J_{2}-J-1)}\nu_{\frac{1}{2}(J_{1}+J_{2}-J-1)}\cdot\cdot\cdot\mu_{J_{1}}\nu_{J_{2}}\alpha}.\]

Now turning to the example, consider the relation of degree two with
spin and parity $2^{-}$: \[
0=\left[\hat{{\cal T}}_{2},\hat{{\cal S}}_{2}\right]_{2}+\frac{i}{3}\sqrt{\frac{7}{2}}\left[\hat{{\cal T}}_{2},\hat{{\cal S}}_{1}\right]_{2}-\frac{2}{3}\sqrt{14}\left\{ \hat{{\cal J}}_{1},\hat{{\cal S}}_{2}\right\} _{2}.\]
According to the above discussion, choose the components of the spin
2 multiplet with highest weight to calculate the prefactors. This
leads to the following covariant relation:

\[
0=-\bra{2,2;2,0}\ket{2;2}\sqrt{\left(\begin{array}{c}
4\\
2\end{array}\right)}\Pi(\mu_{2},\nu_{2})g_{\mu_{1}\nu_{1}}\left[\hat{\Om}^{\mu_{1}\mu_{2}},\hat{\Si}^{\nu_{1}\nu_{2}}\right]\]

\[
+\frac{i}{3}\sqrt{\frac{7}{2}}(i)\bra{2,1;2,0}\ket{2;2}\sqrt{\left(\begin{array}{c}
2\\
2\end{array}\right)}\Pi(\mu_{1},\alpha)\frac{P^{\nu}}{m}\varepsilon_{\nu\hspace{0,1cm}\mu_{2}\nu_{1}}^{\hspace{0,1cm}\alpha}\left[\hat{\Om}^{\mu_{1}\mu_{2}},\hat{\s}^{\nu_{1}}\right]\]

\[
-\frac{2}{3}\sqrt{14}(i)\bra{1,2;1,1}\ket{2;2}\sqrt{\frac{1}{2}\left(\begin{array}{c}
4\\
3\end{array}\right)}\Pi(\mu_{1},\alpha)\frac{P^{\nu}}{m}\varepsilon_{\nu\hspace{0,1cm}\nu_{1}\mu_{2}}^{\hspace{0,1cm}\alpha}\left\{ \hat{w}^{\mu_{1}},\hat{\Si}^{\nu_{1}\nu_{2}}\right\} \]
\[
=-\sqrt{\frac{2}{7}}\sqrt{6}\left[\hat{\Om},\hat{\Si}\right]^{\mu_{1}\mu_{2}}-\frac{1}{3}\sqrt{\frac{7}{2}}\sqrt{\frac{2}{3}}\left[\hat{\Om},\hat{\s}\right]^{\mu_{1}\mu_{2}}+\frac{2}{3}\sqrt{14}\sqrt{\frac{1}{3}}\sqrt{2}i\left\{ \hat{w},\hat{\Si}\right\} ^{\mu_{1}\mu_{2}}.\]

With the definitions introduced above this takes the form:\[
0=\left[\hat{\Om},\hat{\Si}\right]^{\mu_{1}\mu_{2}}+\frac{7}{6}\left[\hat{\Om},\hat{\s}\right]^{\mu_{1}\mu_{2}}-\frac{14}{9}i\left\{ \hat{w},\hat{\Si}\right\} ^{\mu_{1}\mu_{2}}.\]
 One sees that the transformation is done very easily and yields a
one to one mapping without any mixing of the relations. For a more
detailed analysis of the general case see ref.\cite{D.P04}.

\section*{4. Conclusions}

For the manifestly Lorentz covariant presentation of the algebra of
the closed relativistic string we compared two different strategies.

The first method splits the tensor $Q_{\mu\nu\alpha\beta}$ into irreducible
parts i.e. Ricci and Weyl parts. These are linear combinations of
the original rest frame generators. For this reason the covariant
relations of the algebra become linear combinations of the rest frame
relations.

The second method splits the tensor $Q_{\mu\nu\alpha\beta}$ via the
energy-momentum vector into irreducible tensors. These tensors are
covariant extensions of the original rest frame generators. Therefore
they are unique. The relations of the algebra can be made manifestly
covariant separately.

As has been shown in this paper the first method has several disadvantages.
Most importantly, the original generating and induced relations are
mixed. This means that one cannot be sure about the quality of information
in the covariant relation. Further, this mixing means that one has
to calculate more relations than only the generating ones. The total
number of relations increases very fast with the degree of the algebra
so that after the 4th or 5th degree it is impossible to perform all
calculations. Another point is that the covariant generators are not
unique. For a redundance-free presentation this is not desireable.
Last, but not least, gauge degrees of freedom are introduced via the
tetrad (i.e. the arbitrariness of the spacelike part of the basis)
which have to be controlled so that the result is independend of the
choice of the basis.

The second method avoids all these problems. First of all, every generator
has a separate covariant counterpart. Therefore the relations stay
all separate and can be translated one by one. So there is no mixing
and no need to calculate the induced relations. Furthermore, the covariant
generators are as unique as the original ones. And finally, only observable
properties of the string, i.e. the energy-momentum vector, are used
and there is no gauge degree of freedom.

So it is obvious that the formalism which uses the energy-momentum
vector to separate the new generators has great advantages in comparison
with the tetrad formalism.

\hspace{1cm}

\appendix

\section{The Projector on the totally symmetric traceless tensors}

The projector $\Pi(\mu_{1}\cdot\cdot\cdot\mu_{n})$ onto the totally
symmetric traceless tensors in 1+3 dimensions is given as follows:

\[
\Pi(\mu_{1}\cdot\cdot\cdot\mu_{n})A^{\mu_{1}\cdot\cdot\cdot\mu_{N}}=\left(1+\sum_{k=1}^{[\frac{N}{2}]}\Pi_{r=1}^{k}(-1)^{k}\frac{1}{(d+2N-2r-2)}\right.\]

\[
\times\left.\sum_{\mathbb{{P}}(i_{1}\cdot\cdot\cdot i_{N})}\frac{1}{2^{k}k!}\left(\delta^{\nu_{i_{1}}\nu_{i_{2}}}\delta_{\mu_{i_{1}}\mu_{i_{2}}}\times\cdot\cdot\cdot\times\delta^{\nu_{i_{2k-1}}\nu_{i_{2k}}}\delta_{\mu_{i_{2k-1}}\mu_{i_{2k}}}\right)\right)\frac{1}{N!}\sum_{\mathbb{{P}}(1\cdot\cdot\cdot n)}A^{\alpha_{1}\cdot\cdot\cdot\alpha_{N}}\]

where $\mathbb{{P}}(i_{1}\cdot\cdot\cdot i_{N})$ denotes all permutations
of the $N$ indices, 

$\left[\frac{N}{2}\right]=\left\{ \begin{array}{c}
\frac{N}{2}\\
\frac{N-1}{2}\end{array}\right.\begin{array}{c}
\mbox{for N even}\\
\mbox{for N odd}\end{array}$ and $\delta_{\mu_{1}\mu_{2}}:=-\left(g_{\mu_{1}\mu_{2}}-\frac{P_{\mu_{1}}P_{\mu_{2}}}{m^{2}}\right)$
.

The derivation of this formula involves some elementary combinatorics.

\hspace{1cm}

\end{document}